\documentclass[sigconf]{acmart}
\usepackage{booktabs} % For formal tables
\usepackage[utf8x]{inputenc}
\usepackage{threeparttable}
\usepackage{pifont}
\usepackage{xcolor}
\usepackage{bm}
\usepackage{amssymb}
\usepackage{makecell}
\usepackage{hyperref}
\usepackage{paralist}
\usepackage{multirow}
\usepackage{enumitem}

\usepackage{cleveref}
\crefname{algorithm}{Algorithm}{Algorithms}
\crefname{example}{Example}{Examples}
\crefname{part}{Part}{Parts}
\crefname{table}{Table}{Tables}
\crefname{figure}{Figure}{Figures}
\crefname{chapter}{Chapter}{Chapters}
\crefname{section}{Section}{Sections}
\crefname{appendix}{Appendix}{Appendices}

\usepackage{soul}
\usepackage{xcolor}

\newcommand{\ctext}[3][RGB]{%
  \begingroup
  \definecolor{hlcolor}{#1}{#2}\sethlcolor{hlcolor}%
  \hl{#3}%
  \endgroup
}

\newcommand{\feature}[1]{\ctext[RGB]{220,220,220}{\texttt{#1}}}

\usepackage{subcaption}

\usepackage{listings}

\definecolor{codegreen}{rgb}{0,0.6,0}
\definecolor{codegray}{rgb}{0.5,0.5,0.5}
\definecolor{codepurple}{rgb}{0.58,0,0.82}
\definecolor{backcolour}{rgb}{0.95,0.95,0.92}

\lstdefinestyle{mystyle}{
    backgroundcolor=\color{backcolour},
    commentstyle=\color{codegreen},
    keywordstyle=\color{magenta},
    numberstyle=\tiny\color{codegray},
    stringstyle=\color{codepurple},
    basicstyle=\fontsize{7}{9}\ttfamily,
    breakatwhitespace=true,
    breaklines=true,
    captionpos=b,
    keepspaces=true,
    numbersep=5pt,
    showspaces=false,
    showstringspaces=false,
    showtabs=false,
    tabsize=4,
    frame=tblr,
    framerule=0pt,
    columns=flexible
}
\colorlet{punct}{red!60!black}
\definecolor{background}{HTML}{EEEEEE}
\definecolor{delim}{RGB}{20,105,176}
\colorlet{numb}{magenta!60!black}

\lstdefinelanguage{json}{
    basicstyle=\normalfont\ttfamily,
    numbers=left,
    numberstyle=\scriptsize,
    stepnumber=1,
    numbersep=8pt,
    showstringspaces=false,
    breaklines=true,
    frame=lines,
    backgroundcolor=\color{background},
    literate=
     *{0}{{{\color{numb}0}}}{1}
      {1}{{{\color{numb}1}}}{1}
      {2}{{{\color{numb}2}}}{1}
      {3}{{{\color{numb}3}}}{1}
      {4}{{{\color{numb}4}}}{1}
      {5}{{{\color{numb}5}}}{1}
      {6}{{{\color{numb}6}}}{1}
      {7}{{{\color{numb}7}}}{1}
      {8}{{{\color{numb}8}}}{1}
      {9}{{{\color{numb}9}}}{1}
      {:}{{{\color{punct}{:}}}}{1}
      {,}{{{\color{punct}{,}}}}{1}
      {\{}{{{\color{delim}{\{}}}}{1}
      {\}}{{{\color{delim}{\}}}}}{1}
      {[}{{{\color{delim}{[}}}}{1}
      {]}{{{\color{delim}{]}}}}{1},
}

\begin{document}
\title[{Integrating CVL with Multilanguage Annotations for Web Engineering}]{Integrating the Common Variability Language with Multilanguage Annotations for Web Engineering}

\author{Jose-Miguel Horcas}
\affiliation{%
  \institution{Universidad de Málaga, CAOSD Group}
  \city{Málaga}
  \country{Spain}
}
\email{horcas@lcc.uma.es}

\author{Alejandro Cortiñas}
\affiliation{%
  \institution{Universidade da Coruña, Laboratorio de Bases de Datos}
  \city{A Coruña}
  \country{Spain}
}
\email{alejandro.cortinas@udc.es}

\author{Lidia Fuentes}
\affiliation{%
  \institution{Universidad de Málaga, CAOSD Group}
  \city{Málaga}
  \country{Spain}
}
\email{lff@lcc.uma.es}

\author{Miguel R. Luaces}
\affiliation{%
  \institution{Universidade da Coruña, Laboratorio de Bases de Datos\\Enxenio S.L.}
  \city{A Coruña}
  \country{Spain}
}
\email{luaces@udc.es}

\renewcommand{\shortauthors}{Horcas et al.}

\begin{abstract}
Web applications development involves managing a high diversity of files and resources like code, pages or style sheets, implemented in different languages. To deal with the automatic generation of custom-made configurations of web applications, industry usually adopts annotation-based approaches even though the majority of studies encourage the use of composition-based approaches to implement Software Product Lines. Recent work tries to combine both approaches to get the complementary benefits. However, technological companies are reticent to adopt new development paradigms such as feature-oriented programming or aspect-oriented programming. 
Moreover, it is extremely difficult, or even impossible, to apply these programming models to web applications, mainly because of their multilingual nature, since their development involves multiple types of source code (Java, Groovy, JavaScript), templates (HTML, Markdown, XML), style sheet files (CSS and its variants, such as SCSS), and other files (JSON, YML, shell scripts).
We propose to use the Common Variability Language as a composition-based approach and integrate annotations to manage fine grained variability of a Software Product Line for web applications. 
In this paper, we (i) show that existing composition and annotation-based approaches, including some well-known combinations, are not appropriate to model and implement the variability of web applications; and (ii) present a combined approach that effectively integrates annotations into a composition-based approach for web applications. We implement our approach and show its applicability with an industrial real-world system.
\end{abstract}



\copyrightyear{2018}
\acmYear{2018}
\setcopyright{acmcopyright}
\acmConference[SPLC '18]{22nd International Systems and Software Product Line Conference}{September 10--14, 2018}{Gothenburg, Sweden}
\acmBooktitle{SPLC '18: 22nd International Systems and Software Product Line Conference, September 10--14, 2018, Gothenburg, Sweden}
\acmPrice{15.00}
\acmDOI{10.1145/3233027.3233049}
\acmISBN{978-1-4503-6464-5/18/09}


\maketitle

\section{Introduction}
\label{sec:introduction}
Web applications development involves managing commonality and variability spread over  a high diversity of files and resources like code, pages or style sheets, implemented in different languages.
To deal with the automatic generation of custom made configurations of web applications, industry usually adopts annotation-based approaches~\cite{Hunsen2016,Kastner2008CIDE} despite the fact that the majority of studies encourage the use of composition-based approaches~\cite{Bendhun2016,Kastner2008} to implement Software Product Lines (SPLs)~\cite{Apel2013}.
This is mainly because annotations~\cite{Apel2013,Kastner2008CIDE} are simple, flexible, and easy to adopt since they are natively supported by many programming languages. In contrast, composition approaches improve modularization, separation of concerns, and maintenance~\cite{Apel2013}. However, existing composition-based approaches, such as feature-oriented programming (FOP)~\cite{Prehofer1997} or aspect-oriented programming (AOP)~\cite{Kiczales1997}, lack expressiveness and require that industry takes risks and puts high efforts to successfully adopt these new technologies~\cite{Kastner2008CIDE,Kruger2017}.

Moreover, the multilingual nature of web applications, involving multiple types of source code (Java, Groovy, JavaScript), templates (HTML, Markdown, XML), style sheet files (CSS and its variants, such as SCSS), and other kinds of files (JSON, YML, shell scripts), makes extremely difficult, or even impossible to apply some advanced programming models (e.g., FOP, AOP).
Besides, web applications must handle great amounts of fine-grained variability, which can be easily implemented with annotations, but that it would nevertheless be virtually impossible to implement with a composition-based approach. 
Therefore, in a web developing enterprise like Enxenio\footnote{\url{http://www.enxenio.es}}, it was not practical to adopt a new mechanism that did not use annotations and highly difficulties the implementation of the features.
Several works~\cite{Kastner2008,Kruger2016,Behringer2014} try to combine annotative and composition approaches to get their complementary benefits~\cite{Kastner2008}. These works attempt to introduce feature composition into annotation-based approaches~\cite{Kruger2016}, or introduce new implementation layers~\cite{Kastner2008,Behringer2014}, with the goal of bringing composition techniques closer in practice, but as a result they propose complex approaches to be adopted by industry.  
So, our goal is that Enxenio continues using their annotations, but improving the modularity of the code and the traceability between features, variation points, components and final source files.

In this paper, we propose to integrate annotations into a composi\-tion-based approach, contrary to other approaches that extend annotations with composition mechanisms~\cite{Kruger2016,Kruger2017}. Concretely, we use the Common Variability Language (CVL)~\cite{Haugen2008} as a composition-based approach and integrate annotations to manage fine-grained variability of an SPL for web applications. 
We make the following contributions:
\begin{itemize}[noitemsep,nolistsep,leftmargin=\parindent]
\item We show that existing composition and annotation-based approaches, including some well-known combinations~\cite{Kastner2008,Kruger2016}, are not appropriate to model and implement the kind of variability present in web applications.
\item We present a combined approach that effectively integrates annotations into a composition-based approach for web applications.
\item We evaluate our approach and discuss its quality criteria in comparison with classical and combined solutions for implementing SPLs.
\end{itemize}

The rest of the paper is structured as follows. \cref{sec:relatedwork} discusses related work. \cref{sec:motivation} presents our web-based SPL case study and motivates our approach showing the limitations of the existing approaches. \cref{sec:approach} presents our combined approach using CVL. \cref{sec:evaluation} evaluates our approach taking into account different quality criteria and compares it with the existing approaches. Finally, \cref{sec:conclusions} concludes the paper and presents future work.

\vspace*{-0.2cm}
\section{Related Work}
\label{sec:relatedwork}
\begin{figure*}
 \centering
 \includegraphics[width=\linewidth]{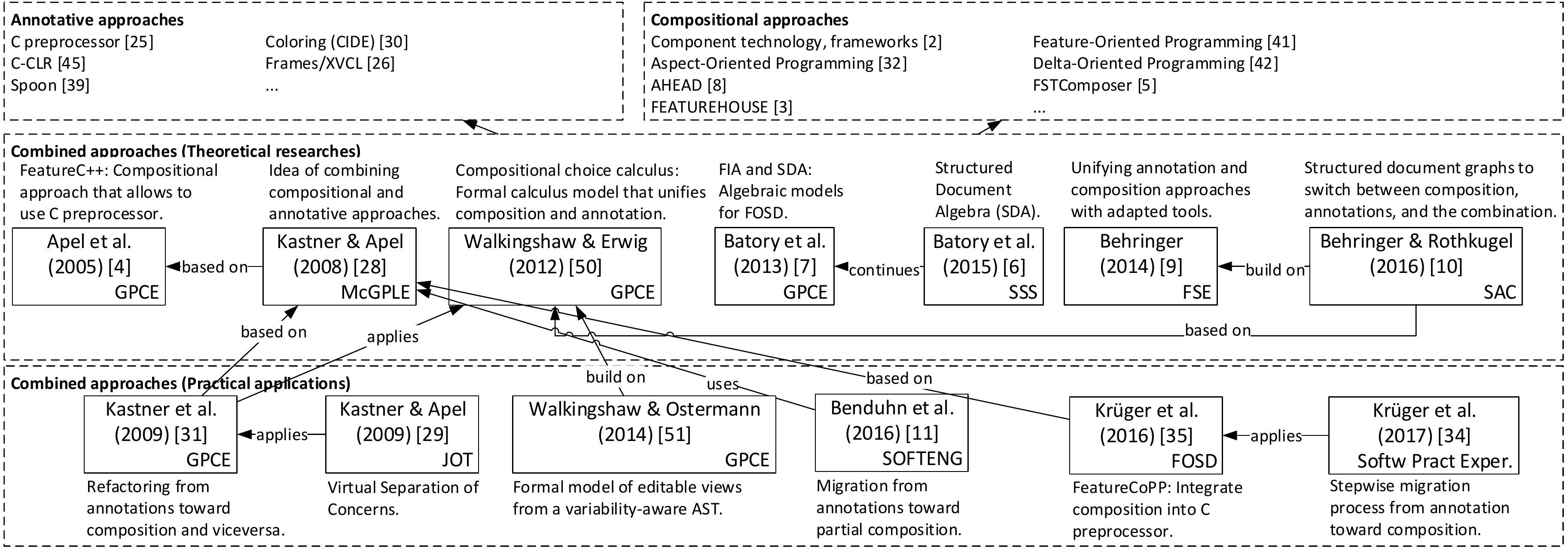}
 \vspace*{-0.7cm}
 \caption{Summary of existing SPL combination techniques.}
\label{fig:relatedwork}
\vspace*{-0.5cm}
\end{figure*}

This section presents related work in the context of the SPL implementation techniques that combines composition and annotative approaches. Figure~\ref{fig:relatedwork} summarizes these works differentiating the theoretical researches and practical applications. 

Although the FeatureC++ approach~\cite{Apel2005} already unintentionally integrated compositional and annotative approaches, by using C preprocessors (\texttt{\#ifdef} annotations)~\cite{Hunsen2016}, it were Kästner and Apel~\cite{Kastner2008} who first formulated the idea of combining both composition and annotative approaches.
Kästner and Apel~\cite{Kastner2008} analyze and compare both composition and annotative approaches, separately in detail, and show the benefits of an integrated approach that introduces an additional implementation layer on top of preprocessors. However, they focus on describing only general ideas for a combined approach, and on discussing the resulting characteristics (granularity, traceability, etc.).

Walkingshaw and Erwig~\cite{Walkingshaw2012}, Batory~\cite{Batory2013,Batory2015}, and Behringer~\cite{Behringer2014,Behringer2016} also provide theoretical researches related with the idea of combining composition and annotative approaches. Walkingshaw and Erwig~\cite{Walkingshaw2012} present \textit{compositional choice calculus}, a formal calculus model to unify composition and annotations, and put it into practice~\cite{Walkingshaw2014} by generating editable documents (views) from a variability-aware \textit{abstract syntac tree}. However, this approach depends on the programming language used. 
Don Batory~\cite{Batory2013,Batory2015} proposes two algebraic models: the \textit{feature interaction algebra} and the \textit{structured document algebra}. 
These models formalize the concept of module with variation points, the composition of them and the decomposition of the modules into smaller parts, simulating annotations for Feature-Oriented Software Development (FOSD)~\cite{Apel2013}. 
Behringer et al.~\cite{Behringer2014,Behringer2016} propose to unify composition and annotative approaches with adapted tools~\cite{Behringer2016}. 
In particular, they propose \textit{structured document graphs}~\cite{Behringer2016} based on the \textit{compositional choice calculus}~\cite{Walkingshaw2012} to change between composition, annotations, and the combination of both approaches in an SPL.

Thenceforth, existing work~\cite{Kastner2009,Kastner2009JOT,Bendhun2016,Kruger2016,Kruger2017} that put into practice the combined approach in SPL are mainly based on the idea of the integrated (hybrid) approach proposed by Kästner and Apel~\cite{Kastner2008}. They claim that the integration is straightforward, conceptually and technically, since it is based on combining existing implementation techniques such as combining preprocessors~\cite{Hunsen2016} or virtual separation of concerns~\cite{Kastner2009JOT} with FOP~\cite{Prehofer1997}, AOP~\cite{Kiczales1997}, or delta-oriented programming~\cite{Schaefer2010}. However, the advantages of these approaches (e.g., modularity, expressiveness,\ldots) mainly depend on the specific composition and annotative implementation techniques used. But the election of the programming model (e.g., using or not using AspectJ) should not be imposed by the SPL implementation mechanism.
In contrast, we propose a composition-based approach with CVL at the architectural and design level, independently of the SPL implementation technique, that integrates the code level annotations within CVL.

All these approaches~\cite{Kastner2009,Kastner2009JOT,Bendhun2016,Kruger2016,Kruger2017} are useful in the scenarios of refactoring annotated SPL in order to utilize, or to migrate toward, composition; and to adopt SPLs from legacy systems (the extractive approach)~\cite{Apel2013}. In particular, Benduhn et al.~\cite{Bendhun2016} apply the integration approach proposed by Kästner and Apel~\cite{Kastner2008} in a real case study, by migrating Berkeley DB from C preprocessors annotations toward partial composition. They demonstrate that although the idea is feasible, the task is challenging, error-prone, and that not all physical separations can be achieved easily.
Krüger et al.~\cite{Kruger2016} present FeatureCoPP (Feature Compositional PreProcessor), an integrated implementation concept that introduces composition into an annotation-based approach. Concretely, they extend the idea of preprocessors to support composition and enable physical separation of concerns similar to FOP. 
In~\cite{Kruger2017}, the migration process from annotation-based toward composition-based approaches is applied to the Berkeley DB case study. In contrast to them~\cite{Kruger2016,Kruger2017}, we propose just the contrary with CVL, that is, we introduce annotations into a composition-based approach.

The conclusion is that technological companies are reluctant to embrace these kinds of combined approaches, mainly when these approaches imposed the adoption of a new programming paradigm. This is even worse if we need to apply a new programming paradigm in the context of web engineering, because we need to use disparate languages and  new languages appear on the market every day.

\section{Motivation and Case Study}
\label{sec:motivation}
Enxenio is a small enterprise that has been producing information systems during the last fifteen years for many different domains, including document management systems, academic management systems or business process support systems. One of their particular field of expertise is geographic information systems (GIS)~\cite{luaces2009, brisaboa2007, places2007}. 
Independently of the domain, most of their clients demand products built on web. Lately, Enxenio, in collaboration with the Databases Laboratory at University of A Coruña, designed and developed an SPL for web-based GIS~\citep{Cortinas2017, Cortinas2017a}, with mostly positive results using an annotation-based approach with their own tooling support. However, it is well-known that maintaining and evolving an SPL following this strategy is an arduous task~\cite{Kastner2008, Spencer1992, Lohmann2006, Schulze2013}, as the number of annotated files grows exponentially and the quality of the annotated source code decreases. 

In this section we describe the particular context of Enxenio and the requirements we took into account in order to choose the most appropriate variability implementation technique. Afterwards, we enumerate the limitations found in existing approaches and tools that made them not suitable to meet our needs. To illustrate our points, we propose a simplified example of a web application that covers most of the needs, regarding variability.

\subsection{Case study}
\label{ssec:casestudy}

Since we need to incorporate a Blog into different web applications, we are interested in defining a  \emph{Blog} SPL. 
A blog is a website where entries (called \emph{post}) are HTML text written by registered users of the blog using a post editor. The blog platform provides different types of editors to write the posts: an HTML, a Markdown, and a WYSIWYG editor. Posts can contain images that can be uploaded or referenced using a URL. Besides having an author and a timestamp, posts can also be linked with specific tags. In order to write a new post, a user needs to authenticate in the web application. The registered users are usually managed by an administrator. We can also allow anonymous users. 
Readers of the blog can comment on the posts, and we can even decide if they need to be registered users in order to comment or any anonymous user can do that. The blog can also have one or more widgets in the front page to manage the tags, comments, or files; and the user interface, including the administration pages, can be internationalized. Finally, as a mean to debug the code properly, we can choose to add some extra logging in the code (i.e., a logger).

\subsection{Requirements}
\label{ssec:requirements}
Web development nowadays involves a high number of different technologies and languages. HTML, CSS and JavaScript are the three standard languages that can be interpreted by any web browser rendering a webpage. However, developers usually use additional programming languages, for instance server-side web technologies such as Java, Python or Ruby, to implement the web application functionality (e.g., connect to a database).
Concretely, the code of our Blog SPL involves using a total of 12 different languages or file types, each one including some variable parts: (i) Java and JavaScript programming languages; (ii) some markup languages such as HTML, XML and Markdown; (iii) some style sheet languages such as CSS and SCSS; (iv) some data serialization languages such as YAML and JSON; and, (v) some others standard file formats such as property files or script files. Therefore, it is required that our SPL solution can handle this set of languages, but also any other new language that may appear in the future. 
\vspace{-0.05cm}
\begin{description}
\item[Req1.] \emph{The approach needs to be independent of the language, that is, a multilingual SPL approach.}
\end{description}
\vspace{-0.05cm}
As previously discussed, web applications expose a high degree of fine-grained variability that usually affects most of the web artifacts specified in different languages. For instance, in the development of web applications it is very common to use many third-party libraries. Each library has its own particular way of being deployed and used, and it may involve, for example, adding some lines in HTML, or defining a parameter in a configuration file. In this case, annotating the lines that vary would be more efficient and easy to understand than modeling this variability with a component.
\begin{description}
\item[Req2.] \emph{The approach needs to support fine-grained variability.}
\end{description}

In order to evolve and maintain the source code of our SPL, it is desirable to maintain each piece of code implementing a feature, separately from the rest of code. Also, we would like to easily identify and relate any feature with the final code that implements it, so that the developer can easily add or modify features or code variants maintaining the consistency of the SPL. 
Indeed, this is the main motivation of our work, to improve the evolution management and maintenance of very large and complex SPLs. 
\begin{description}
\item[Req3.] \emph{The approach must keep the feature traceability between all layers of our SPL and preserve the separation of concerns.}
\end{description}

In the majority of application domains, the product automatically generated using an SPL approach does not need to be modified before its market release. In the web domain, this is not possible, because there are some client customizations that must be performed manually, because it is not possible to model them as part of the SPL. One common example is the adaptation of the user interface to the corporative style of the customer company (i.e., logo, colors, messages style,\ldots). Even if we use one of the well-known Content Management System (CMS)~\cite{Lima2014} such as \textit{Wordpress}, developers need to modify the actual code file of the templates to adapt the style of the view. So, the  product code generated by the SPL normally needs some modifications for a specific customer, and thus, a  requirement for our Blog SPL is:
\begin{description}
\item[Req4.] \emph{The generated code should be as clear and simple as possible, without including any additional glue code or artificial mechanisms that affects the code legibility.}
\end{description}

Besides the technical points, one problematic issue when trying to change the methodology of an enterprise to an SPL-based one is the reticence of the development teams~\cite{Berger2013a} and the costs in time to train both the platform engineers and the product engineers.
Recently, 
in Enxenio, we have tried to totally change the development procedures in order to improve the productivity of the company, but with no success. Developers only adopted a minor part of the new methodology that was 
partially abandoned. 
With that bad experience in mind, one of the conditions expressed by the manager team in order to adopt the SPL technology for their products was:
\begin{description}
\item[Req5.] \emph{To keep the development procedures flexible.}
\end{description}
This requirement particularly means:
\begin{inparaenum}[1)]
    \item avoiding technological changes only justified as a mean to fit the implementation technique; 
    \item tooling support independent of the development IDE and of the operating system;
    \item keeping the flexibility in the development of the different components; and
    \item lower specific formation requirements, since most of the employees at Enxenio are newly graduated, and their formation is focused on the development procedures and the improvement of their programming skills.
\end{inparaenum}

\subsection{Limitations of existing approaches}
\label{ssec:limitations}
Existing approaches, both compositional and annotative approaches, as well as the combined approaches do not completely fulfill the requirements of web engineering~\cite{Kastner2008,Kruger2016,Bendhun2016}.

Composition-based approaches do not support fine-grained variability and the generated code is usually full with code implementing artificial methods created to handle the variability and that make the code very hard to understand and maintain. 
Besides that, composition-based approaches are not able to work independently of the language (e.g., HTML, CSS, Ruby, JSON,\ldots). There are some attempts to make a multi-language approaches such as FeatureHouse~\cite{Apel2013a}, but the fact is that every different language requires to introduce a new plugin supporting the new language specificities, increasing the complexity of the product generation, especially when several languages are used in the same product. Even more, if the syntax of a programming language changes, something very common in web development, the plugin stops working until it is re-programmed to support the new syntax.
As an example, most of Java 8 artifacts are still unsupported by AHEAD~\cite{Batory2004} or FeatureHouse~\cite{Apel2013a}. So, Enxenio cannot use those approaches that have to be maintained by third parties.

Annotation-based approaches can work totally independently of the language as text preprocessors.  
However, tooling support for annotations are not designed to be used specifically within an SPL approach, and they do not bear in mind the separation of concerns principle. This means that the traceability of features and artifact code is not easy. 
Moreover, it is well known that the maintenance of the platform code using annotations is a nightmare~\cite{Apel2013}. 

The existing combined approaches that mixes both compositional and annotative approaches have similar drawbacks. This is the case of the generic combination approach~\cite{Kastner2008}, that depends on the actual composition and annotation-based approaches used. Besides that, introducing composition into an annotation-based approach, like in FeatureCoPP~\cite{Kruger2016}, suffers the same problem regarding the poor quality of the generated code, modified with artificial methods created only to handle variability. Moreover, how well traceability and separation of concerns is achieved depends on the implemented solution and not on the approach itself (e.g., FOP).

The approach presented in this paper aims to overcome these limitations and also to address the requirements of SPLs for web applications commented above. In Section~\ref{sec:evaluation} we discuss, in detail, the quality criteria of existing approaches in comparison with our approach. Next section presents our approach.

\vspace*{-0.2cm}
\section{Our Approach: Integrating composition and annotations in CVL}
\label{sec:approach}
This section details our combined approach for composition and annotation variability with the Common Variability Language (CVL). First, we present how CVL works for composition. Second, we introduce our multilingual annotations. Then, we integrate the annotations in the CVL approach. Finally, we apply the approach to our Blog product line.

\begin{figure}
 \centering
 \includegraphics[width=\linewidth]{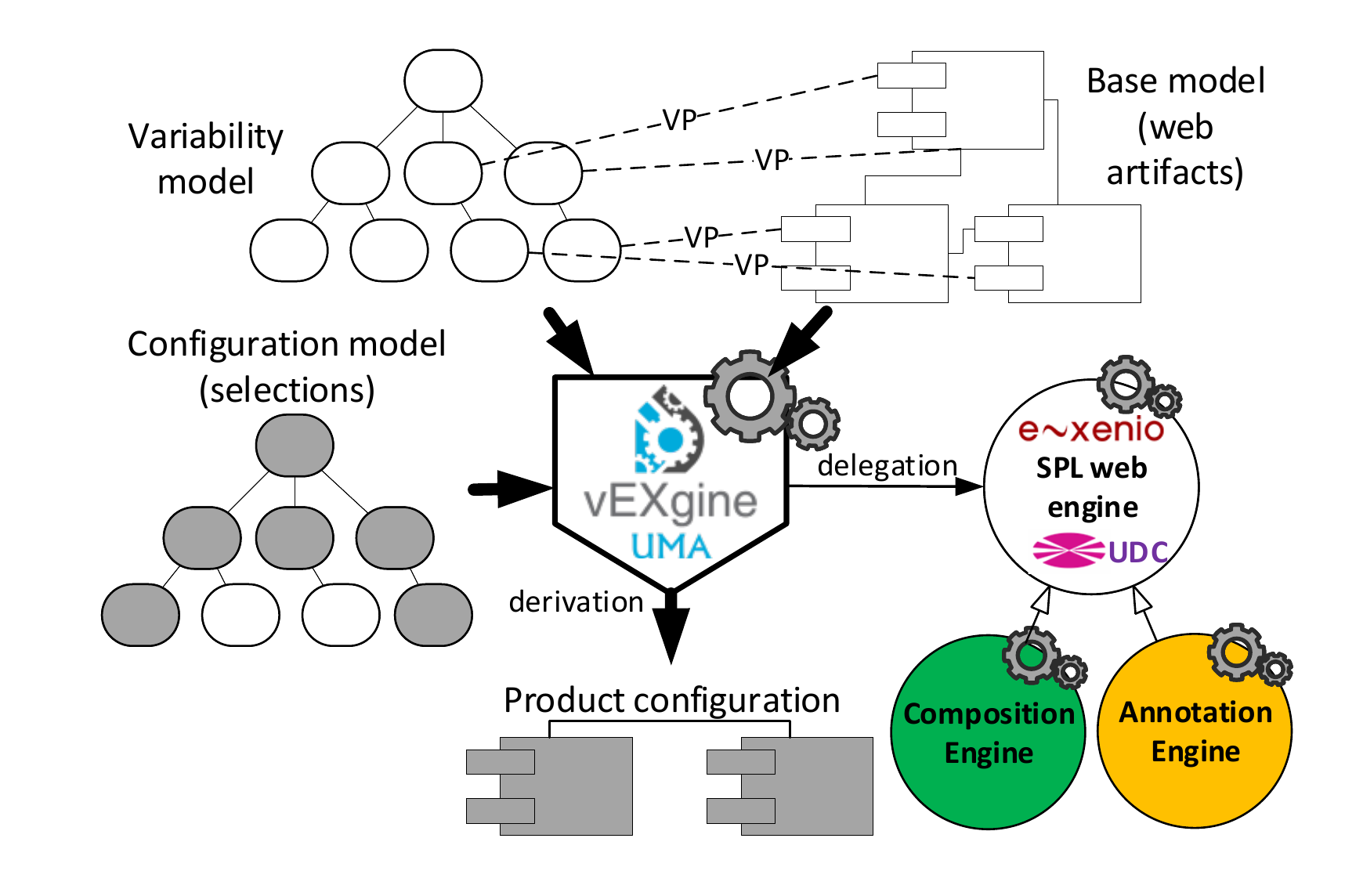}
 \vspace*{-0.8cm}
 \caption{Our approach based on CVL and its tool support.}
\label{fig:approach}
\vspace*{-0.6cm}
\end{figure}

As shown in \cref{fig:approach}, CVL specifies, in separate models, the variability that can be applied to a \textit{base model}. The base model is a model in the domain language that can be defined using a MOF-based metamodel (e.g., UML) and normally does not contain any information about variability. In the context of web engineering the base model represents all the artifacts that are part of the SPL of web applications, from code artifacts to templates or style files. 
CVL specifies the variability information in a separate variability model, similar to traditional feature models~\cite{Kang1990}. The variability model also defines  
the points of the base model that are variable and can be modified during the derivation process --- i.e., the variation points (VPs).
A configuration model (i.e., a selection of features) describes how the variability is resolved to produce a configured product from the base model. CVL relies on its executable engine to automatically derive a product with the variability resolved. 

\begin{table*}
	\caption{Compositional and annotative variation points.}
	\label{tab:variationpoints}
    \centering
    \scriptsize
    \vspace*{-0.4cm}
    \begin{threeparttable}
\begin{tabular}{clcl}
\toprule
& Variation Point & Type & Semantic\\
\midrule
\multirow{6}{*}{Predefined\tnote{1}} & ObjectExistence & Compositional & It indicates the existence of an object in the base model.\\
& LinkExistence & Compositional & It indicates the existence of a link in the base model.\\
& ObjectSubstitution & Compositional & It substitutes an object with another one in the final product model.\\
& FragmentSubstitution & Compositional & It substitutes a set of objects and links (a fragment) with another fragment in the base model.\\
& ParametricSlotAssignment & Compositional & It assigns a new value to a variable.\\
& \ldots & \ldots & \ldots \\
\midrule
\multirow{4}{*}{Custom} & OVP (Existence) & Annotative & It indicates the existence of portions of text in the files related to an existing object of the base model. \\
& OVP (Assignment) & Annotative & It assigns a new value to an annotated variable in the files related to an existing object of the base model.\\
& OVP (Uses) & Annotative & In indicates the existence of a portion of code implementing a <<uses>> link between two components for different features\\
& \ldots & \ldots & \ldots \\
\bottomrule
\end{tabular}
        \begin{tablenotes}
            \item[1] The complete taxonomy of variation points predefined in CVL is available in~\cite{Haugen2012CVL}. 
        \end{tablenotes}
	\end{threeparttable}
        \vspace*{-0.4cm}
\end{table*}

We propose to extend the CVL approach to allow specifying not only coarse-grained variability, but fine-grained variability that can be resolved with annotations.
To bring the CVL approach in practice we use \textit{vEXgine}~\cite{Horcas2017vexgine}, a customizable and extensible implementation of the CVL execution engine that fully supports the derivation process, including a delegation mechanism that can be extended with different delegation engines. In particular, for this paper, we extend vEXgine to delegate the variability resolution process in a \textit{scaffolding-based derivation engine} adapted from~\cite{Cortinas2017a} to support the separation of the source code into different components, being each of them composed itself by a set of artifacts. 
In order to make the integration easier and flexible, we built a web service that receives all the input data from vEXgine, such as the base models and the semantic of the variation points for a specific variability configuration, and generates the source code of the final product. 
We call this tool \textit{SPL Web Engine}, and it is in charge of resolving the variability of different granularity: (i) coarse-grained and medium-grained variability for those features that can be implemented following a compositional approach (\texttt{Composition Engine}); and (ii) fine-grained variability for those features that need to be implemented following an annotative approach (\texttt{Annotation Engine}).

\subsection{The composition-based approach of CVL}

The CVL approach is, by nature, an orthogonal composition-based approach since elements of the base model can be composed, removed, substituted, etc. through the CVL variation points. Variation points specify how the elements of the base models are modified by defining specific modifications to be applied by means of model-to-model (M2M) transformations. The semantics of these transformations are specific to the kind of each variation point. During CVL’s execution, the CVL engine (vEXgine in our approach) delegates its control to an M2M engine in charge of executing the transformations specific of each variation point. In our approach, vEXgine delegates its control to a more generic engine (the SPL Web Engine) to execute the semantic of the variation points. Only the semantic of those variation points bound to a selected feature in a configuration model will be executed during variability resolution.

Some of the variation points supported and predefined by CVL for composition are the existence of elements of the base model (\textit{ObjectExistence}), the links between them (\textit{LinkExistence}), the assignment of an attribute's value (\textit{ParametricSlotAssignment}), or the replacement of a set of elements with another set of elements (\textit{FragmentSubstitution}), among others (upper part of \cref{tab:variationpoints}). A very powerful variation point is the Opaque Variation Point (OVP) that enables to define a custom-made variation point with a new tailored semantic, not pre-defined in CVL.

\subsection{Multilanguage annotations for web SPLs}
\label{ssec:multilanguageannotations}
In this subsection we will present the annotations used in our combined approach.
As previously said, in CVL the base model does not contain any information about coarse-grained variability --- i.e., this kind of variability is specified in the variability model with the variation points.  However, in the context of our SPL for web applications, we do allow that  artifacts of the base model contain variability information, but fine-grained variability, implemented by means of annotations. These annotations correspond with the variability that makes no sense to be managed by a composition-based approach (e.g., to model variable parts of a HTML web form).

\begin{figure*}[tbp]
\centering
\begin{subfigure}[t]{.5\textwidth}
\begin{lstlisting}[language=HTML,numbers=left]
<div class="btn-group">
    <button type="button" class="btn btn-default" ng-click="$ctrl.toggleHTML()" title="Toggle HTML / Markdown">
        <span class="glyphicon glyphicon-open" aria-hidden="true"></span>
    </button>
    <!--% if (feature.imageFromURL) { %-->
    <button type="button" class="btn btn-default" ng-click="$ctrl.insertImageFromURL()" title="Insert from URL">
        <span class="glyphicon glyphicon-picture"></span>
    </button>
    <!--% } %-->
    <!--% if (feature.imageUploading) { %-->
    <button type="button" class="btn btn-default" ng-click="$ctrl.uploadImage()" title="Upload image">
        <span class="glyphicon glyphicon-open"></span>
    </button>
    <!--% } %-->
    <button type="button" class="btn btn-default" ng-click="$ctrl.showMarkdownHelp()" title="Show Markdown help">
        <span class="glyphicon glyphicon-open"></span>
    </button>
</div>
\end{lstlisting}
  \vspace*{-0.2cm}
  \caption{Example of annotated HTML code.}
  \label{fig:html-code}
\end{subfigure}%
\begin{subfigure}[t]{.5\textwidth}
\begin{lstlisting}[language=java,numbers=left]
angular.module('blog').component('editor', {
    bindings : { post: '=' },
    controller: controller,
    templateUrl: 'app/components/editor/editor.html'
});
function controller($showdown, /*% if (feature.imageUploading) { */FileUploader, /*% } */ModalService) {

    this.toggleHTML = function() { ... };
    
    /*% if (feature.imageFromURL) { */
    this.insertImageFromURL = function() {
        // implementation of the feature
    };
    /*% } */
    /*% if (feature.imageUploading) { */
    this.uploadImage = function() {
        // implementation of the feature
    };
    /*% } */
    
    this.showMarkdownHelp = function() { ... };
}
\end{lstlisting}
  \vspace*{-0.2cm}
  \caption{Example of annotated JavaScript code.}
  \label{fig:js-code}
\end{subfigure}
\vspace*{-0.4cm}
\caption{Example of multilingual annotations for two artifacts in different languages.}
\label{fig:annotated}
\vspace*{-0.5cm}
\end{figure*}

The annotation engine used by the SPL Web Engine processes each file as plain text. This means that our annotations are language independent or multilanguage since we can annotate any text-based web artifact used in our case study (HTML, CSS, Java, etc.). 
When an annotation is found, the code associated with this annotation is evaluated. 
% written always in JavaScript, 
This code is different depending on the kind of variation point (lower part of \cref{tab:variationpoints}), and can define: (i) the existence of annotations in files related to elements of the base model (\textit{Existence}); (ii) the assignment of values through annotations to elements of the base model (\textit{Assignment}); and, (iii) the interaction between two elements from the base model (\textit{Uses}). The  annotations are simple JavaScript code embedded in comments (see in Figure~\ref{fig:annotated}, \texttt{if} sentences for the \textit{Existence} variation point).
In~\cref{fig:annotated} we can see simplified excerpts of annotated code for the functionality of the blog associated with the inclusion of images in a post (a) for the view (HTML), and (b) for the controller (JavaScript). In this example, all annotations implement an \textit{Existence} variation point. In the JavaScript code you can also note that if the feature selected is \textit{imageUploading} the annotation adds a new parameter \textit{FileUploader}, which is not present if the other feature \textit{imageFromURL} is selected. This is a good representative example of how fine-grained annotations can be.

Note that the annotations are comments in both of the excerpts shown, and they will not interfere with the source code editors. This is a very simple but really interesting feature of our derivation engine: it allows customizing the delimiters for the annotations depending on the file. This is not done with any extra plugin, but linking each file extension with a particular delimiter. Developers would prefer to use the syntax of the comments that are native in every language. This way, they can work with their favorite IDE and tools without having to deal with intrusive annotations that make the code not compile or that will be marked as syntax errors by for example a HTML editor.

\subsection{Integrating fine-grained variability in CVL}
CVL is intentionally a compositional approach that is applied at a high level of abstraction like the architectural level instead of working at code level. However the CVL approach is unaware of how the features are physically separated into code units implementing the base model (e.g., components, aspects, feature modules), and therefore, any composition-based approach at the code level could be applied, such for instance, FOP or AOP.
Assuming features are separated the best possible in code units, an annotative approach can be used to additionally annotate code units when the variability affects finer levels~\cite{Kastner2008}. So, one code unit can implement a variable feature and at the same time contain variable code text.
On the other hand, in web applications sometimes it is not possible to physically separate independent features  in different code units, as desirable. Or a developer reluctant to adopting our new approach prefer to use annotations to implement variable features at code level.  In these cases, we can use annotations to identify parts of the code implementing different variable features. 
To handle all these case we propose to introduce annotations (our multilingual annotations) within the CVL approach to be able to handle the fine-grained variability from a high abstract level. This allows CVL to also resolve the variability defined by the annotations, apart from the compositional variability, during the product derivation.

\begin{figure}
 \centering
 \includegraphics[width=\linewidth]{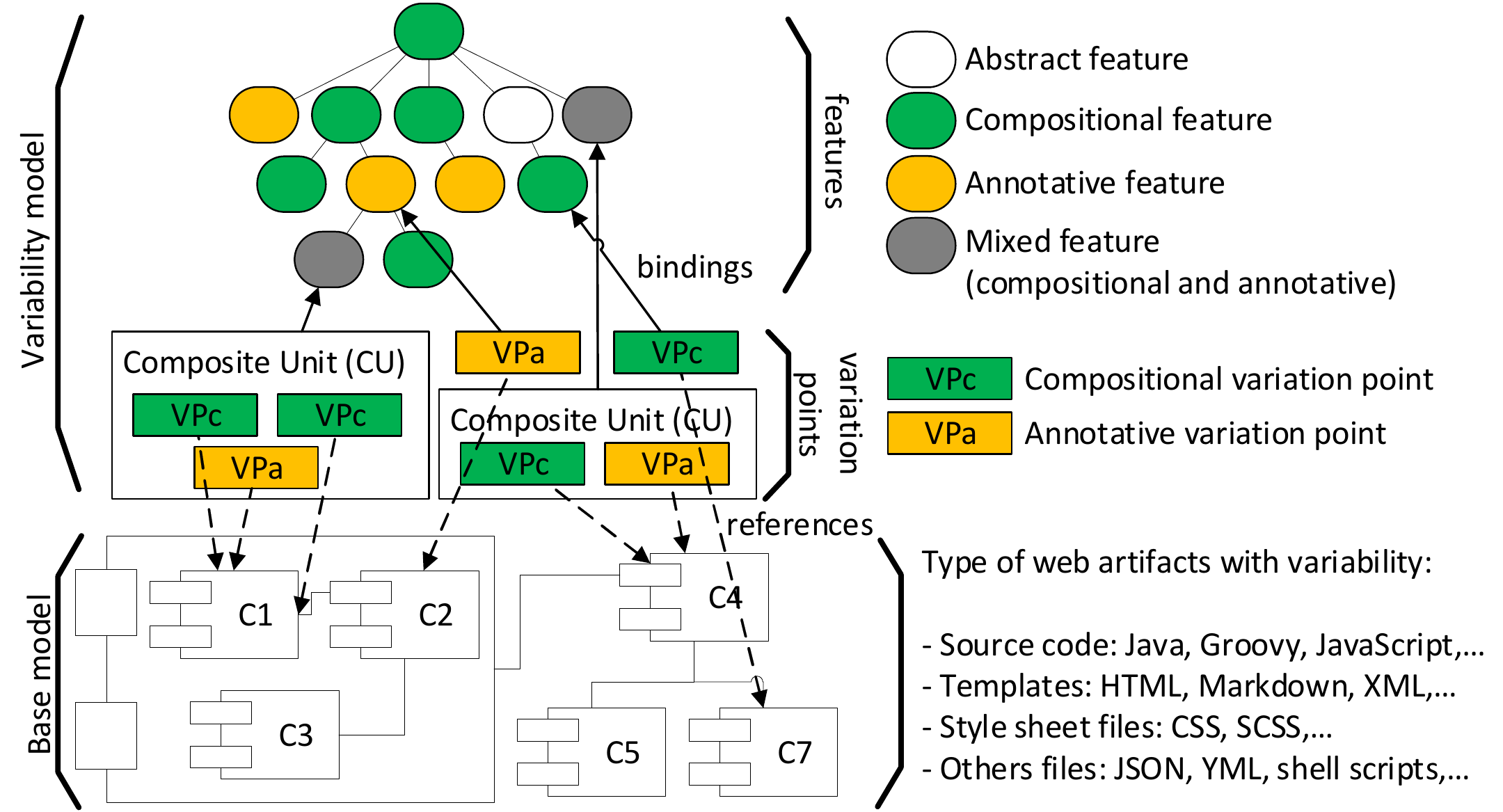}
 \vspace*{-0.6cm}
 \caption{Integrating composition and annotations in CVL.}
\label{fig:schema}
\vspace*{-0.4cm}
\end{figure}

Figure~\ref{fig:schema} shows an schema of our combined approach with CVL.
The variability model consists of two main parts: (1) an abstract level with the feature tree (i.e., VSpec tree in CVL terminology) as in traditional feature models~\cite{Kang1990}, and (2) a concrete level with the variation points (VPs).
At the concrete level (i.e., the variation points), it is possible to distinguish the variability granularity that affects each component since variation points refer directly artifacts of the base model that implement each feature.
We classify variation points into two categories:
\begin{description}[noitemsep,nolistsep,leftmargin=0.5\parindent]
\item[Compositional variation point (VPc).] Compositional variation points define the coarse-grained variability that is applied at the architectural level. These are the traditional variation points provided by CVL such as \textit{ObjectExistence}, \textit{LinkExistence}, \textit{FragmentSubstitution}, or \textit{ParametricsSlotAssignment}. The semantic of the VPc is predefined by CVL, but we can also define our own semantic for VPc using Opaque Variation Points (OVPs). 

\item[Annotative variation point (VPa).] Annotative variation 
points define the fine-grained variability that is applied at a lower level of abstraction --- i.e., at the artifact level such as source code or web templates. An VPa is an Opaque Variation Point (OVP) the semantic of which generically specifies that ``there is an annotation bound to the selected feature and the annotation is located in the component/artifact this variation point refers to''.
\end{description}

Lower part of \Cref{tab:variationpoints} shows some annotative variation points we define to handle fine-grained variability. From this classification of variation points we can now distinguish the granularity of each feature. First,  mandatory features are always present in every generated product, so there is not variability to handle. Secondly, it is possible to include abstract features when necessary for organization purpose (e.g., group optional features), but since they do not represent a concrete variable artifact, they are not bound to any variation point. 
Finally, variable features can be implemented as compositional, annotative, or both (mixed), based on the nature of the feature and on the analysis of the developer. A compositional feature is bound to a compositional variation point, and its semantic will be resolved by the \texttt{Composition Engine} (see Figure~\ref{fig:approach}). An annotative feature is bound to an annotative variation point, which will be resolved by the \texttt{Annotation Engine}.
A mixed feature (i.e., a feature that involves composition and annotations) is bound to a set of variation points encapsulated in a \textit{Composite Unit} (\textit{CU}). A CU in CVL is a unit of modularization that contains a set of variation points to jointly handle the variability of a specific feature or set of features. 
If the mixed feature is selected in a configuration the semantic of all variation points belonging to the bound CU will be processed by the SPL web engine. 

Note that it is also possible to encapsulate in a CU, many variation points of the same type (compositional or annotative). This is useful in those cases where the feature implementation is scattered because of a bad design. So, thanks to the use of CVL, our approach can improve the feature traceability of the SPL independently from the implementation technique used at the code level.

\begin{figure*}[t]
 \centering
 \includegraphics[width=\linewidth,height=0.6\textheight]{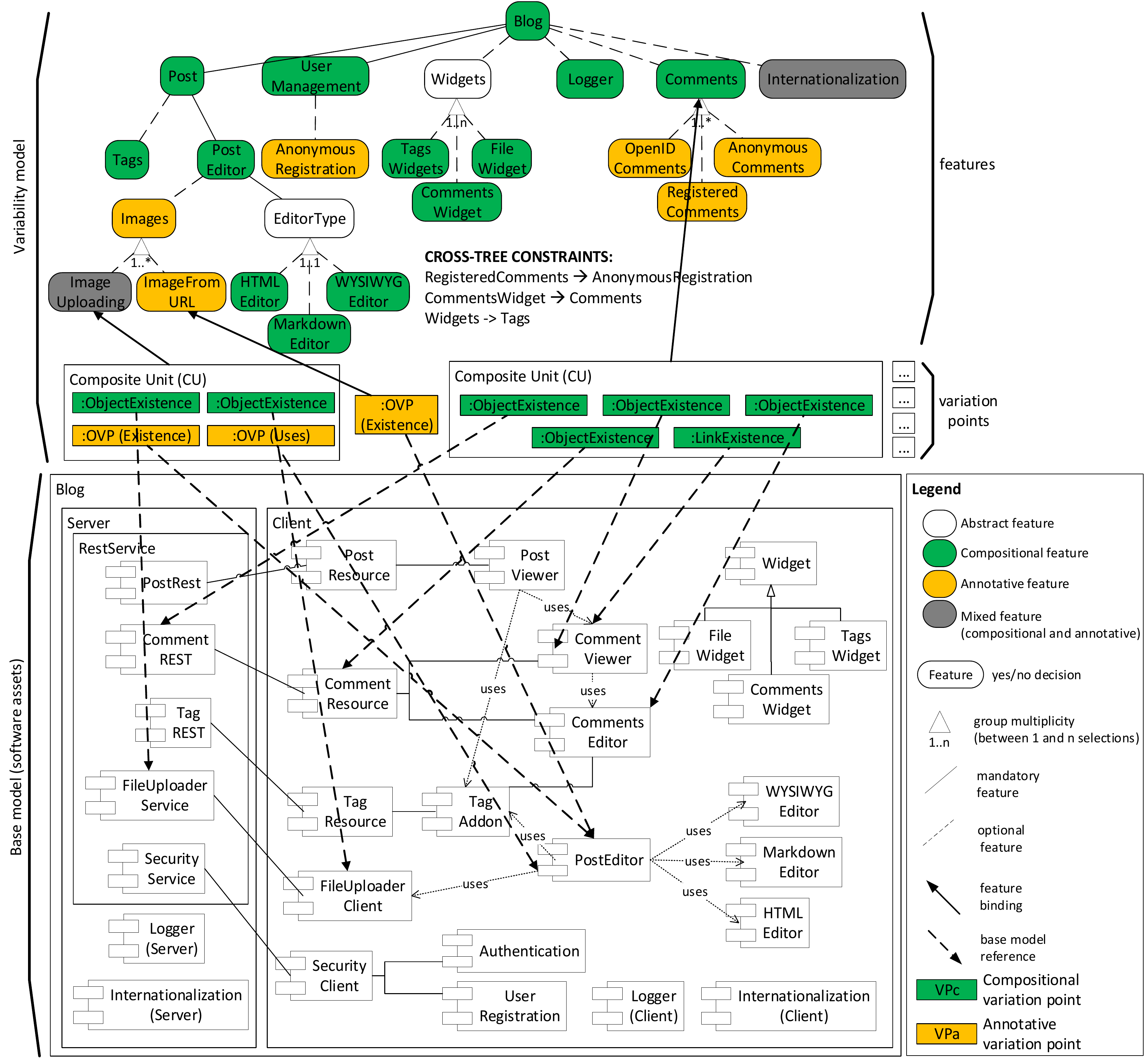}
 \vspace*{-0.6cm}
 \caption{Our approach applied to the Blog product line case study.}
\label{fig:example}
\vspace*{-0.4cm}
\end{figure*}

\subsection{Applying our approach to the Blog SPL}
\cref{fig:example} shows how our approach is applied to the Blog SPL. 
It presents the variability model specified in CVL that includes the two kinds of variation points (compositional and annotative), and the references to the variable artifacts of the base model. Note that in order to simplify the figure we have omitted some links connecting features, variation points and base model.

On the one hand, there are some compositional features that can be grouped in a CU. This is the case of the feature \feature{Comments}, which adds several modules into the final product as specified by the links of each variation point. Each of these variation points is linked to one or several components handling the API of the comments in the server side of the architecture (\texttt{CommentREST}), and some client-side components for writing a comment (\texttt{CommentEditor}), viewing existing comments (\texttt{CommentViewer}) and handling the communication with the API of the comments (\texttt{CommentResource}).
When the feature \feature{Comments} is selected in a configuration, the final product will include all these modules providing the functionality of managing the comments, while if the feature \feature{Comments} is not present in a configuration, all the related modules will be excluded from the final product.

On the other hand, there are some annotative features that require a finer degree of variability such as the \feature{ImageFromURL} feature (see~\cref{fig:annotated}). The annotative variation point associated with this feature includes a reference to the component affected by the annotation.
When the \feature{ImageFromURL} feature is selected in a configuration, the derivation engine will resolve the annotation with the semantic specified and will generate the final product's code with the variability resolved and without any annotations left (in~\ref{fig:annotated} (a) lines 5, 9-14 are removed and also in (b) comments of line 9, 13, 17 and lines 19-23). The HTML and Javascript code show respectively one variant of the view and the controller implementations of the \texttt{PostEditor} component (implemented using the MVC pattern).

Finally, there are also some mixed features (compositional and annotative) whose implementation is much simpler using annotations but which also requires the inclusion of a component as for example the \feature{ImageUploading} and the \feature{Internationalization} features. For instance, the feature \feature{ImageUploading} can be implemented with barely a few lines added to the JavaScript source code of the different editors (see~\cref{fig:annotated}). However, it also requires a generic component able to handle the file uploading in both client (\texttt{FileUploaderClient}) and server side (\texttt{FileUploaderService}). All the variability of the \feature{ImageUploading} feature is encapsulated in a CU, and thus, when this feature is selected in a configuration, all variation points will be applied together. Note that the order in which the variability is resolved does not affect the final product, but it impacts the performance of the derivation process. The derivation engine first resolves the compositional variation points, and then the annotative variation points. This way we prevent to resolve fine-grained variability of components that will be not present in the final product. 
\section{Evaluation}
\label{sec:evaluation}
This section discusses and compares our approach to the pure composition and annotation-based approaches and to the most relevant integrated approaches~\cite{Kastner2008,Kruger2016}.  
We partly base the discussion on the quality criteria for SPL implementation techniques defined in~\cite{Apel2013} (\textit{feature traceability}, \textit{separation of concerns}, \textit{information hiding}, \textit{granularity}, \textit{uniformity}, and \textit{preplanning effort}) but from a different point of view, that is, the architectural level where CVL works. In addition, we also incorporate others interesting quality criteria that are recommendable for SPL implementation, such as the support for multiple \textit{languages}, the \textit{variability type} supported, \textit{automation}, \textit{maintainability}, \textit{evolution}, and \textit{ tool support}. We also compare our approach with a previous solution for web applications used by Enxenio, based only on annotations (Section~\ref{sec:motivation}). To do that, we quantitatively assess those quality criteria that can be measured with a metric in a Web-based GIS product line (around 45K LOC) developed by Enxenio~\citep{Cortinas2017, Cortinas2017a}.

\begin{description}[noitemsep,nolistsep,leftmargin=0.5\parindent]
\item[Feature traceability.] Feature traceability describes the mapping between a feature in the variability model and its implementation in an artifact or set of artifacts. For compositional approaches, this mapping depends on the implementation technique. It is said~\cite{Apel2013,Kruger2016} that the mapping is direct as the artifact that implements a feature can be traced to a single code unit (component, module, aspect,\ldots). However, this `direct' trace is implicitly done by name conventions since the only way to identify and relate the feature in the variability model and the artifact that implements that feature is using the same identifiers for the feature and the artifact, and/or using dedicated tools~\cite{Kastner2008CIDE,Heidenreich2008}. 
Moreover, in annotative approaches, feature traceability is poorly supported because annotations can be scattered over multiple artifacts, and traceability in this case is usually a matter of tool support~\cite{Apel2013}. Traceability in combined approaches is weaker than in pure compositional approaches because existing integrating approaches~\cite{Kastner2008,Kruger2016} are basically annotative approaches that try to introduce composition.

With CVL, our approach allows tracing a feature (compositional or annotative) explicitly from the variability model to the artifacts. This is done thanks to the bindings and references defined by the variation points (see Figure~\ref{fig:schema}) that explicitly bind each feature of the variability model to the target artifacts in the software architecture. Although an annotative feature is scattered in multiple artifacts due to a bad design, our approach allows explicitly identifying the artifacts affected by the annotation.

\item[Separation of concerns.] Separation of concerns refers to the ability to separate feature functionality into cohesive implementations~\cite{Apel2013}, even when features are crosscutting concerns like the \feature{Logger} and the \feature{Internationalization} features in our example. Separation of concerns depends on the implementation carried out by the developers~\cite{Kruger2016}, that in turns depends on the programming paradigm used (e.g., FOP, AOP). 
For most composition-based approaches, separation of concerns is intended, but not for annotation-based approaches~\cite{Kastner2008} in which this separation can be simulated with tool support (e.g., CIDE~\cite{Kastner2008CIDE}).

Likewise in composition-based approaches, separation of concerns in our approach also depends on the implementation of developers. However, in contrast to existing combined approaches, to understand the variability of a feature modeling a concern, it is not necessary to look at the artifact implementation because the variability model explicitly exposes the variability information through the variation points.

\item[Information hiding.] Information hiding is the separation of a module into internal and external part (e.g., an interface).
While some composition-based approaches such as frameworks or components technology provide good support for information hiding, other compositional approaches such as FOP or AOP do not~\cite{Apel2013}. Annotation-based approaches prevent information hiding because of the fine-grained nature of the features~\cite{Apel2013}. Information hiding in combined approaches depends on the composition mechanism used, but normally is weaker than in pure compositional approaches~\cite{Kruger2016}.

Our approach supports information hiding well due to the software architecture vision. For compositional features, we assume their functionality is encapsulated in generic modules or artifacts (not necessary in a one-to-one relation) that we can modify, delete, or replace by other modules that implement the same interfaces, but we are unaware of the specific implementation technique (e.g., AOP). So a module of the base model could be a JavaScript component, an aspect in AspectJ, or even a web template in HTML. Note that, from the point of view of the information hiding, our approach works as the CORE approach~\cite{Alam2013,Schottle2016} for composition. For annotative features, our approach hides, at the architectural level, the internal variability of the modules and explicitly indicates which modules are affected by fine-grained variability. In any case, our approach does not support information hiding when dealing with variability at the code level as in the majority of annotation-based approaches~\cite{Kruger2016}.

\item[Granularity.] Granularity describes the level on which variability is implemented~\cite{Apel2013}. Compositional approaches only provide coarse-grained or medium-grained variability at the level of components, classes, methods extensions, etc., while annotative approaches support well fine-grained variability  at the level of statement, parameters, or expressions~\cite{Kastner2008CIDE}.

Similarly to other integrated approaches~\cite{Kastner2008,Kruger2016}, our approach supports all levels of granularity. On the one hand, multilingual annotations support fine-grained variability at the most finer statements, being possible to annotate the same code line with different annotations, that is, our approach does not enforce undisciplined annotations~\cite{Apel2013}. On the other hand, the composition mechanism provided by CVL allows coarse-grained variability on top of the hierarchical structure (e.g., packets, directories,\ldots) where application modules are physically stored.

\item[Uniformity.] Uniformity refers to the principle that all artifacts (annotated or composed) should be encoded and synthesized in a similar manner or style, regardless the implementation technique~\cite{Apel2013}. Both pure compositional and annotative approaches often enforce a common style (e.g., preprocessors for annotations or aspects for AOP). But combined approaches~\cite{Kastner2008,Kruger2016} enables developers to use different styles at the same time.  

Our approach allows representing all artifacts subject to variation as software components of an architectural model (e.g., in UML). The variability of both annotated and composed artifacts are indistinguishable without the information contained in the variation points.

\item[Preplanning effort.] SPL engineering always incurs a certain a\-mount of preplanning~\cite{Apel2013}. 
While compositional approaches usually require substantial preplanning activities, annotation-based approaches allows introduce annotations to artifacts with lower efforts~\cite{Apel2013}, as occurs also for combined approaches based mainly on annotations~\cite{Kastner2008,Kruger2016}. Our approach requires to build the variability model (similar effort that for the feature model) in addition to the specification of the variation points, their bindings to the features and their references to the artifacts; resulting in a high amount of preplanning. However, note that the specification of the variability model including the variation points is done only once at the domain engineering stage of the SPL engineering process.

\item[Multilanguage and language independence.]

Most of the composition based approaches are monolingual, that  is, they work exclusively for one language. Particularly, most of them work on Java, such as AHEAD~\cite{Batory2004}, AspectJ~\cite{Kiczales1997}, or DeltaJ~\cite{Koscielny2014DeltaJ}, among others.
There are exceptions such as FeatureHouse~\cite{Apel2013a}, based on FSTComposer~\cite{Apel2008}, that support more than one language, but a plugin or extension is needed for each language. 

Although, there are also some annotation-based tools that only work for a specific language (e.g., Spoon~\cite{Pawlak2005} for Java), annotative approaches are usually multilanguage such as C preprocessor, CIDE~\cite{Kastner2008CIDE}, Frames/XVCL~\cite{Jarzabek2003}, or C-CLR~\cite{Singh2007}, as well as the main commercial alternatives such as Gears\footnote{\url{http://www.biglever.com/}} or pure::variant\footnote{\url{http://www.pure-systems.com/}}. 

Regarding some of the existing alternatives that combine composition and annotations, in the generic combination~\cite{Kastner2008} the approach itself is independent of the language but it relies on the particular engines used for composition and annotation. For example, on FeatureC~\cite{Kruger2016} they rely on FeatureHouse~\cite{Apel2013a} and on C preprocessor, so they had to develop a FeatureHouse plugin to support C preprocessor annotations on feature-based modules.
FeatureCoPP~\cite{Kruger2016} is based solely on C preprocessor and therefore is independent of the language. However, there is still no tooling support to test this approach with a multilingual product line and evaluate how intrusive the annotations are. 

\begin{table*}[t]
	\caption{Quality criteria of our approach in comparison with existing SPL implementation strategies.}
	\label{tab:quality-criteria}
    \vspace*{-0.4cm}
    \centering
    \scriptsize
	\begin{threeparttable}
		\begin{tabular}{lccccc}
			\toprule
            Quality Criteria & Composition-based approaches & Annotation-based approaches & Generic combination~\cite{Kastner2008} & FeatureCoPP~\cite{Kruger2016} & \textbf{Our Approach} \\
            \midrule
            Feature traceability & naming & tool support & naming and tool support & naming and tool support & bindings and references \\
            Separation of concerns & intended & simulated with tool support & implementation dependent & implementation dependent & implementation dependent \\
            Information hiding & comp. mechanism dependent & prevented & comp. mechanism dependent & comp. mechanism dependent & independent\tnote{AL}\\
            Granularity & coarse- and medium-grained & fine-grained & all levels & all levels & all levels\\
            Uniformity & common style & common style & different styles & different styles & common style\tnote{AL} \\
            Preplanning effort & high & low & low & low & medium \\
            \makecell{Language independence\\and multilanguage} & \makecell{host language dependent,\\ monolingual} & language independent & approach dependent & language independent & \makecell{language independent,\\ multilingual} \\
            Variability type & 
            	\makecell{positive, function,\\ plat./env.} & 
                negative, optional & 
                \makecell{positive, negative,\\ optional, alternative,\\ function, plat./env.} & 
                \makecell{positive, negative,\\ optional, alternative,\\ function, plat./env.} & 
                \makecell{positive, negative,\\ optional, alternative,\\ function, plat./env.,\\ and custom}\\
            Automation & comp. mechanism dependent & tool support & \makecell{comp. mechanism dependent\\ and tool support} & \makecell{comp. mechanism dependent\\ and tool support} & tool support \\
            Maintainability & low & high & medium & medium & medium \\
            Evolution & manual & manual & manual & manual & automatic with algorithms \\
            Tool Support & IDE dependent & C preprocessor, CIDE,\ldots & Not dedicated tool & Not dedicated tool & vEXgine~\cite{Horcas2017vexgine}, and Enxenio tool~\cite{Cortinas2017a}\\
			\bottomrule
		\end{tabular}
        \begin{tablenotes}
        	\item[AL] At the architectural level.
        \end{tablenotes}
	\end{threeparttable}
    \vspace*{-0.4cm}
\end{table*}
\begin{table}
	\caption{SPL for web-based GIS applications.}
    \label{tab:metrics}
	\centering
    \scriptsize
    \vspace*{-0.4cm}
\begin{tabular}{lrr}
\toprule
Metric & Annotative approach & CVL combined approach \\
\midrule
\#features & 128 & 128\\
\#products (configurations) &  255,704 & 255,704 \\
\#artifacts (code files, templates,\ldots) & 621 & 621\\
\midrule
\#annotated artifacts & 504 & 166 \\
\#annotations & 2,592 & 1,694 \\
\bottomrule
\end{tabular}
\vspace*{-0.6cm}
\end{table}

Our approach is completely language independent. For composition, we model the variability of the product line using UML components that can be composed themselves by any kind of artifacts, regardless of the language they are written. For annotations, our SPL Web Service in charge of resolving the variability does not need any kind of adapter or plugin, thanks to our multilanguage annotations (\cref{ssec:multilanguageannotations}).

\item[Variability type.] An SPL approach supports different types of variability categorized as~\cite{Gacek2001}: positive (functionality is added), negative (functionality is removed), optional (code is included), alternative (code is replaced), function (functionality changes), and platform/environment changes.
Compositional approaches often support positive, function, and platform/environment variability. Annotative approaches usually support negative and optional variability.
Combined approaches based on annotations~\cite{Kastner2008,Kruger2016} can support also alternative variability.

Our approach supports all type of variability thanks to the predefined variation points of CVL (Table~\ref{tab:variationpoints}), but also allows to incorporate new types of variabilities user-defined with the OVPs.

\item[Degree of automation and effort.] The degree of automation is a measure that compares the software elements (e.g., number of components, lines of code) that are manually defined with those that are automatically generated~\cite{Horcas2016}.
This metric allows discussing about the development effort, degree of reuse and productivity of applying a specific approach. We define the degree of automation of our approach as the comparison between the number of elements (i.e., code artifacts, templates, files,\ldots) automatically generated when the variability is resolved ($\#e_a$) and the number of elements manually defined ($\#e_m$) in order to manually resolve the variability of a specific product:

\vspace*{-0.3cm}
\begin{equation}
\label{eq:degreeAutomation}
	\text{Degree of Automation} = \frac{\#e_a}{\#e_a+\#e_m}
\end{equation}

As shown in~\cref{tab:metrics}, we observe that the web-based GIS product line contains 621 artifacts, where 504 are variable or contain some degree of variability. This means that to manually generate a specific product we have to manually modify up to 504 different artifacts. Whereas applying an SPL approach as the presented in this paper we automatically resolve the variability of those artifacts, obtaining a degree of automation of 81.16\%.
However, the same degree of automation is achieved by any other SPL approach, so this metric makes sense when considering the developers' efforts when applying and maintaining an specific SPL.

\item[Maintainability.] Maintainability is the ease with which a software product can be modified. Maintenance in SPL is more complex because changes in a module can affect various products. Here we are interested in the maintainability of the SPL instead of the individual generated product after delivery. So, the main goal is to evaluate the impact of maintaining the artifacts of the features that compose the SPL~\cite{Vale2015}.

We observe in~\cref{tab:metrics} that in our previous approach, developers needs to manage 2592 annotations scattered among 504 artifacts. In contrast, in our CVL approach, annotations are reduced up to 1694 (35\% less annotations) scattered among 166 artifacts, reducing the annotated artifacts to be managed up to 67\%. This large reduction in the number of annotated artifacts is due to the fact that many annotations affect complete files to handle coarse-grained variability, and with the new approach these annotations have been modeled in CVL as compositional variation points.

\item[Evolution.] Evolution is the ability to modify the SPL to support changes at the domain engineering level, as for example to incorporate new features or functionalities to the SPL. Normally, independently of the SPL approach, this is a manual task to be performed.
In contrast, the SPL built on our approach can be automatically evolved applying some evolution algorithms formally defined in~\cite{Horcas2016EDOC}. These algorithms allow incorporating changes to the CVL variability model and propagating those changes to the configurations and generated products.

\item[Tool support.] SPL approaches are viable and useful to the extent that they are supported by appropriate tools. 
Composition-based approaches are supported by tools that depend on the implementation mechanisms such as FeatureIDE~\cite{Thum2014FeatureIDE} for FOP, or AspectJ~\cite{Kiczales1997} or AspectC/C++~\cite{Coady2001AspectC,Spinczyk2005AspectCPP} for AOP. These tools are often language and IDE dependent. Most of annotation-based approaches are supported by C preprocessor, but there is also some specific tool to manage annotations such as CIDE~\cite{Kastner2008CIDE}. Combined approaches have not a dedicated tool and the developer usually needs one or more tools to support both compositional and annotative approaches. Our approach is supported by the vEXgine tool~\cite{Horcas2017vexgine} and the SPL Web Engine of Enxenio~\cite{Cortinas2017a} that work in conjunction as explained in Section~\ref{sec:approach}.

\end{description}

\Cref{tab:quality-criteria} summarizes and compares the results of our approach with pure composition-based and pure annotation-based approaches and with the two more well-known integration approaches of Kästner and Apel~\cite{Kastner2008} and FeatureCoPP~\cite{Kruger2016}.

\section{Conclusions and Future Work}
\label{sec:conclusions}
We have presented an integrated solution that combines annotations into the composition-based approach of CVL to handle the variability existing in web applications at different levels of granularity. To do so, we have extended the CVL approach and provided appropriate tool support.

Our approach keeps the mapping between the features and their implementation artifacts through the bindings and references from the variation points to the feature model and to the base models, respectively. So we provide uniform and good support for feature traceability for both compositional and annotative strategies, at the architectural level. The proposed solution is completely language independent through the use of MOF-based models at the architectural level for coarse-grained variability and by means of multilanguage annotations for fine-grained variability.

As future work, we plan to evaluate our approach with several SPLs in the web engineering context. In parallel, pursuing a better maintenance support, we need to add analytic features to our tools (e.g., looking for \emph{dead features} or checking consistency of annotations in code linked to OVPs in CVL).

\section*{Acknowledgements}
{
\footnotesize The work of the authors from the Universidad de Málaga is supported by the projects Magic P12-TIC1814 and HADAS TIN2015-64841-R (co-financed by FEDER funds). 
The work of the authors from the Universidade da Coruña has been funded by Xunta de Galicia/FEDER-UE CSI: ED431G/01; GRC: ED431C 2017/58. MINECO-AEI/FEDER-UE Datos 4.0: TIN2016-78011-C4-1-R; Flatcity: TIN2016-77158-C4-3-R. EU H2020 MSCA RISE BIRDS: 690941.
}

\bibliographystyle{ACM-Reference-Format}
\bibliography{references}

\end{document}